\documentclass[preprint,showpacs,preprintnumbers,amsmath,amssymb,floatfix,endfloats*]{revtex4}
\usepackage{graphicx}% Include figure files
\usepackage{dcolumn}% Align table columns on decimal point
\usepackage{bm}% bold math
\usepackage{amsfonts}

\DeclareMathAlphabet{\mathsfsl}{OT1}{cmr}{bx}{it}
\begin{document}
%----------------------------------------------------------------------%
% Title
%----------------------------------------------------------------------%
\title{Reversible plastic events during oscillatory deformation of amorphous solids}
\author{Nikolai V. Priezjev}
\affiliation{Department of Mechanical and Materials Engineering,
Wright State University, Dayton, OH 45435}
\date{\today}
\begin{abstract}

The effect of oscillatory shear strain on nonaffine rearrangements
of individual particles in a three-dimensional binary glass is
investigated using molecular dynamics simulations.  The amorphous
material is represented by the Kob-Andersen mixture at the
temperature well below the glass transition.   We find that during
periodic shear deformation of the material, some particles undergo
reversible nonaffine displacements with amplitudes that are
approximately power-law distributed.   Our simulations show that
particles with large amplitudes of nonaffine displacement exhibit a
collective behavior; namely, they tend to aggregate into relatively
compact clusters that become comparable with the system size near
the yield strain.   Along with reversible displacements there exist
a number of irreversible ones.   With increasing strain amplitude,
the probability of irreversible displacements during one cycle
increases, which leads to permanent structural relaxation of the
material.

\end{abstract}


% \pacs{61.43.Fs, 61.43.-j, 62.20.F-, 83.10.Rs}

%   64.70.pj Polymers
%   81.05.Kf, Glasses (including metallic glasses)
%   61.43.Fs Glasses
%   64.70.P-  Time-dependent properties; relaxation (for glass transitions)
%   61.43.-j  Disordered solids, structure
%   66.30.Pa  Diffusion in nanoscale solids
%   62.20.F-  Deformation: mechanical properties of solids
%   83.10.Rs  Computer simulation of molecular and particle dynamics

\maketitle

\section{Introduction}

Understanding the structure-property relationship of amorphous
polymers and metallic glasses is important for many technological
and biomedical applications~\cite{Pelletier14,Voigtmann14}.   The
mechanism of plastic deformation in amorphous materials involves an
accumulation of highly localized structural rearrangements of atoms
or the so-called shear transformation
zones~\cite{Argon79,Falk98,Weitz07}. It was recently demonstrated
that in sheared glasses a strong correlation exists between a
collective rearrangement of small groups of atoms and
quasi-localized soft modes, or ``soft spots", which are analogous to
dislocations in crystalline
solids~\cite{Tanguy10,Manning11,Schoenholz14,MaFalk14,Tanguy15}.
Molecular dynamics (MD) simulations suggest that specific atomic
packing configurations with the most unfavorable local coordination
polyhedra are more likely to participate in soft spots in metallic
glasses~\cite{Ma12,MaFalk14}.     Furthermore, a local plastic event
in sheared amorphous solids induces long-range deformation that in
turn might trigger secondary events and give rise to
avalanches~\cite{Lemaitre09}.    In related studies, it was shown
that a local reversible shear transformation in a quiescent system
results in cage jumps (discrete events where particles escape from
cages of their neighbors) whose density is larger in the cases of
weakly damped dynamics or slow shear
transformation~\cite{Priezjev15,Priezjev15a}.

\vskip 0.05in

In recent years, the mechanical response of amorphous materials to
cyclic shear was examined
experimentally~\cite{Dennin08,Biroli09,Losert12,Arratia13,Spaepen14,Cipelletti14,Arratia14,Echoes14,Ganapathy14},
by means of molecular dynamics
simulations~\cite{Priezjev13,Sastry13,Reichhardt13,HernHoy13,Priezjev14,Losert14,Kawasaki15,IdoNature15}
and continuum modeling~\cite{Perchikov14}. It was shown that at
strain amplitudes below a critical value, particle trajectories are
reversible after either one or several cycles and the diffusion is
suppressed~\cite{Dennin08,Sastry13,Priezjev13,Reichhardt13,HernHoy13,Arratia14,Losert14}.
In contrast, at larger strain amplitudes, the number of cage
breaking events increases and the particle dynamics becomes
spatially and temporally
heterogeneous~\cite{Biroli09,Priezjev13,Sastry13,Priezjev14,Losert14}.
However, the nature of the transition (a sharp crossover versus a
continuous non-equilibrium phase
transition~\cite{Cipelletti14,Ganapathy14,Kawasaki15}) and its
relation to chaotic behavior~\cite{Reichhardt13} remain not fully
understood.

\vskip 0.05in

In the previous MD studies of binary~\cite{Priezjev13} and
polymeric~\cite{Priezjev14} glasses under oscillatory shear strain,
the structural relaxation was studied by analyzing the self-overlap
order parameter and the dynamic susceptibility.     It was found
that at small strain amplitudes, the system dynamics is nearly
reversible during several thousands of oscillation periods. On the
other hand, at strain amplitudes above the critical value, the
memory of the initial state is lost during several cycles and a
large fraction of particles undergo irreversible
displacements~\cite{Priezjev13,Priezjev14}. Remarkably, it was shown
that at the critical strain amplitude that separates slow and fast
relaxation dynamics, the number of particles involved in a
correlated motion reaches maximum~\cite{Priezjev13,Priezjev14}.
Furthermore, the cage-breaking events were identified from a
sequence of particle positions at the end of each cycle and studied
at different strain amplitudes.   In particular, it was demonstrated
that dynamic facilitation of mobile particles by their neighbors
becomes increasingly pronounced at larger strain
amplitudes~\cite{Priezjev13,Priezjev14}.    However, the analysis of
particle displacements in the previous studies did not include
plastic rearrangements that occur during a single oscillation cycle.

\vskip 0.05in

In this paper, the structural relaxation process in a binary glass
under cyclic loading is investigated using molecular dynamics
simulations.      We find that while the system dynamics at small
strain amplitudes is reversible after each cycle, most particle
undergo nonaffine displacements with amplitudes that are broadly
distributed.    Moreover, the results of numerical simulations
indicate that large nonaffine rearrangements are spatially
heterogeneous, with the typical length scale on the order of the
system size near the yield strain.

\vskip 0.05in

The rest of the paper is structured as follows.   The molecular
dynamics simulation model and the deformation protocol are described
in the next section. The analysis of nonaffine displacements of
particles at different strain amplitudes is presented in
Sec.\,\ref{sec:Results}.   Brief conclusions are provided in the
final section.

\section{Details of molecular dynamics simulations}
\label{sec:MD_Model}

\begin{figure}[t]
\includegraphics[width=8.5cm,angle=0]{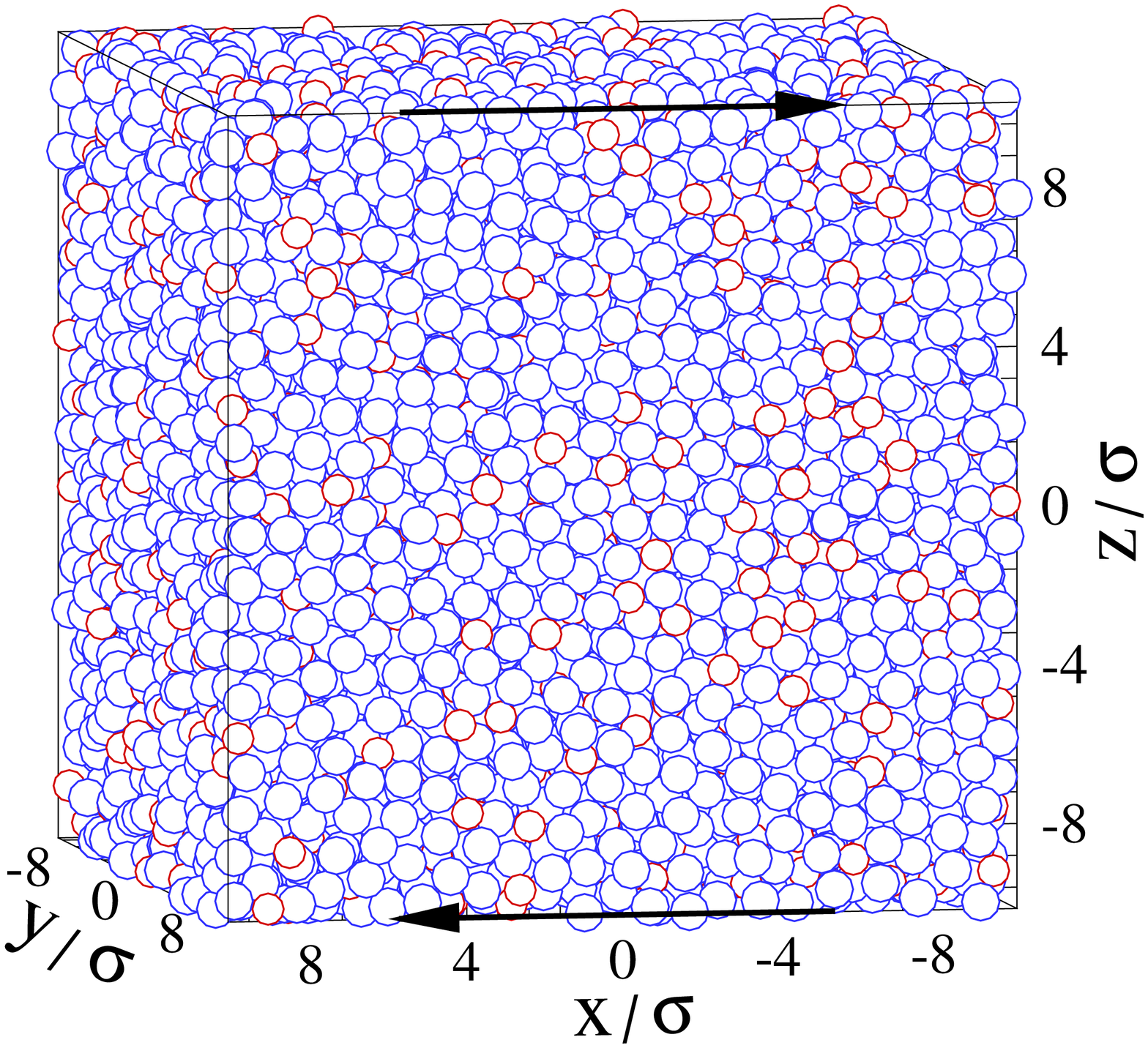}
\caption{(Color online) A snapshot of an equilibrated binary LJ
glass that is subject to oscillatory shear strain in the $xz$ plane
(indicated by solid arrows).  Atoms of type A are denoted by large
blue circles and type B by small red circles. The Lees-Edwards
periodic boundary conditions are applied in the $xz$ plane. }
\label{fig:snapshot_system}
\end{figure}

We perform molecular dynamics simulations of a binary (80:20)
Lennard-Jones (LJ) glass model, which was first introduced by Kob
and Andersen~\cite{KobAnd95}.    The three-dimensional system
consists of $N=10\,000$ particles placed in a periodic box as
illustrated in Fig.\,\ref{fig:snapshot_system}.     In this model,
any two particles $\alpha,\beta=A,B$ interact via the LJ potential,
which is defined as
\begin{equation}
V_{\alpha\beta}(r)=4\,\varepsilon_{\alpha\beta}\,\Big[\Big(\frac{\sigma_{\alpha\beta}}{r}\Big)^{12}\!-
\Big(\frac{\sigma_{\alpha\beta}}{r}\Big)^{6}\,\Big],
\label{Eq:LJ_KA}
\end{equation}
where the parameters are set as $\varepsilon_{AA}=1.0$,
$\varepsilon_{AB}=1.5$, $\varepsilon_{BB}=0.5$, $\sigma_{AB}=0.8$,
$\sigma_{BB}=0.88$, and $m_{A}=m_{B}$~\cite{KobAnd95}.   The cutoff
radius is $r_{c,\,\alpha\beta}=2.245\,\sigma_{\alpha\beta}$.  In
what follows, the units of length, mass, energy, and time are set
$\sigma=\sigma_{AA}$, $m=m_{A}$, $\varepsilon=\varepsilon_{AA}$, and
$\tau=\sigma\sqrt{m/\varepsilon}$, respectively.    The simulations
were carried out at a constant density
$\rho=\rho_{A}+\rho_{B}=1.2\,\sigma^{-3}$ in a cubic box of linear
size $L=20.27\,\sigma$.   The Newtons equations of motion were
solved numerically using the fifth-order Gear predictor-corrector
integration scheme~\cite{Allen87} with a time step $\triangle
t_{MD}=0.005\,\tau$.

\vskip 0.05in

% deformation protocol

The system was initially equilibrated in the absence of shear at the
temperature $1.1\,\varepsilon/k_B$, which is well above the glass
transition temperature
$T_g\approx0.45\,\varepsilon/k_B$~\cite{KobAnd95}. Here $k_B$ is the
Boltzmann constant.    Then, the temperature was gradually reduced
with the rate of $10^{-5}\,\varepsilon/k_B\tau$ to the final
temperature $T=10^{-2}\,\varepsilon/k_B$.   After additional
$5\times10^6$ MD steps, the periodic shear strain was applied in the
$xz$ plane according to
\begin{equation}
\gamma(t)=\gamma_{0}\,\,\textrm{sin}(\omega t),
\label{Eq:strain}
\end{equation}
where $\omega$ is the oscillation frequency and $\gamma_{0}$ is the
strain amplitude.    The simulations were carried out with the
oscillation frequency $\omega\tau=0.001$ and, correspondingly, the
period $T=2\pi/\omega=6283.19\,\tau$.   In our study, the shear
deformation was implemented using the SLLOD algorithm~\cite{Evans92}
combined with the Lees-Edwards periodic boundary conditions in the
$xz$ plane.    The constant temperature of the system was maintained
by rescaling the $\hat{y}$ component of the velocity for each
particle.    In addition, periodic boundary conditions were applied
along the $\hat{y}$ direction (perpendicular to the plane of shear).
After discarding several cycles, the positions of all particles were
saved every $T/12=523.60\,\tau$ during fifty oscillation periods.
The data were accumulated in eight independent samples for each
strain amplitude.

\section{Results}
\label{sec:Results}

% definition of D2min

The local plastic event in sheared amorphous materials can be
detected by computing nonaffine displacements of particles with
respect to its neighbors~\cite{Falk98}.   We first evaluate the
transformation matrix $\mathbf{J}_i$ that best maps all bonds
between a particle $i$ and its nearest neighbors at times $t$ and
$t+\Delta t$.  The nonaffine displacement of the particle $i$ is
computed as follows:
\begin{equation}
D^2(t, \Delta t)=\frac{1}{N_i}\sum_{j=1}^{N_i}\Big\{
\mathbf{r}_{j}(t+\Delta t)-\mathbf{r}_{i}(t+\Delta t)-\mathbf{J}_i
\big[ \mathbf{r}_{j}(t) - \mathbf{r}_{i}(t)    \big] \Big\}^2,
\label{Eq:D2min}
\end{equation}
where the sum is taken over the neighboring atoms within the radius
$r_c=1.5\,\sigma$ from the position vector
$\mathbf{r}_{i}(t)$~\cite{Ma12,MaFalk14}.  In our study, the time
lag $\Delta t$ was varied in the range from $T/12$ to $50\,T$.  Note
also that the oscillation period was chosen so that the time
interval between consecutive particle configurations,
$T/12=523.60\,\tau$, is much larger than the typical timescale of
irreversible rearrangements of particles~\cite{Falk98}.

% comment on sensitivity on number of neighbors

\vskip 0.05in

% distribution of D2min during all possible intervals during 10 T

\begin{figure}[t]
\includegraphics[width=12.cm,angle=0]{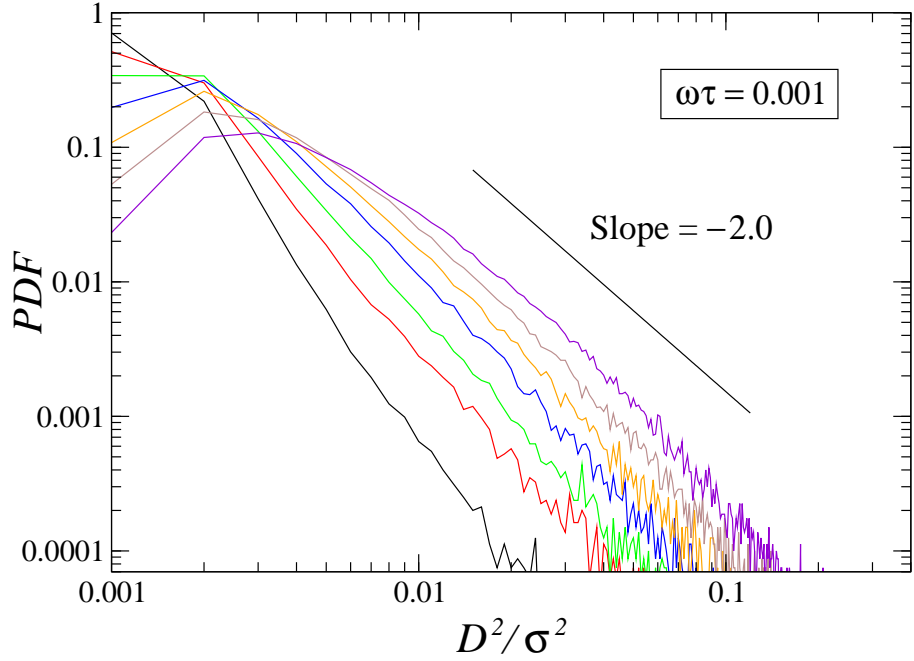}
\caption{(Color online) The normalized probability distribution
function of $D^2(t, \Delta t)$ defined by Eq.\,(\ref{Eq:D2min}) for
the strain amplitudes
$\gamma_{0}=0.01,~0.02,~0.03,~0.04,~0.05,~0.06$, and $0.07$ (from
left to right).  The quantity $D^2(t, \Delta t)$ was measured at
$t=0$ and $\Delta t = T/4$, where $T$ is the oscillation period. The
straight line with the slope $-2$ is plotted for reference. }
\label{fig:d2min_pdf}
\end{figure}

The probability distribution function of the quantity $D^2$ for
different strain amplitudes is presented in
Fig.\,\ref{fig:d2min_pdf}. The data were collected in bins
$0.001\,\sigma^2$ at the time lag $\Delta t=T/4$ and time $t=0$. It
can be observed in Fig.\,\ref{fig:d2min_pdf} that at strain
amplitudes $\gamma_0\leqslant0.07$, most of the particles undergo
affine displacements characterized by small values
$D^2\lesssim0.004\,\sigma^2$.    At the same time, however, the
displacement of some particles deviates significantly from a linear
strain field and the quantity $D^2$ becomes relatively large.   The
tails of the probability distribution functions are approximately
power-law distributed with a slope of the decay approaching $-2$ at
the strain amplitude $\gamma_0=0.07$.  Thus, the threshold for the
local plastic deformation is somewhat arbitrary; and in what
follows, we define it to be $D^2=0.01\,\sigma^2$.   Similarly, a
small fraction of the atomic size was chosen as a critical value of
$D$ in the previous MD studies of actively deformed metallic
glasses~\cite{Peng11,MaFalk14}.    Note also that the definition of
$D^2$ in Eq.\,(\ref{Eq:D2min}) involves averaging over changes in
the nearest-neighbor distances; and, therefore, the threshold
$D^2=0.01\,\sigma^2$ is consistent with the value $0.1\,\sigma$ used
to define cage jumps in the previous studies on oscillatory shear
deformation of binary glasses~\cite{Biroli09,Priezjev13,Priezjev15}.

\vskip 0.05in

% examples of nonaffine rearrangements at gamma0=0.05

\begin{figure}[t]
\includegraphics[width=12.cm,angle=0]{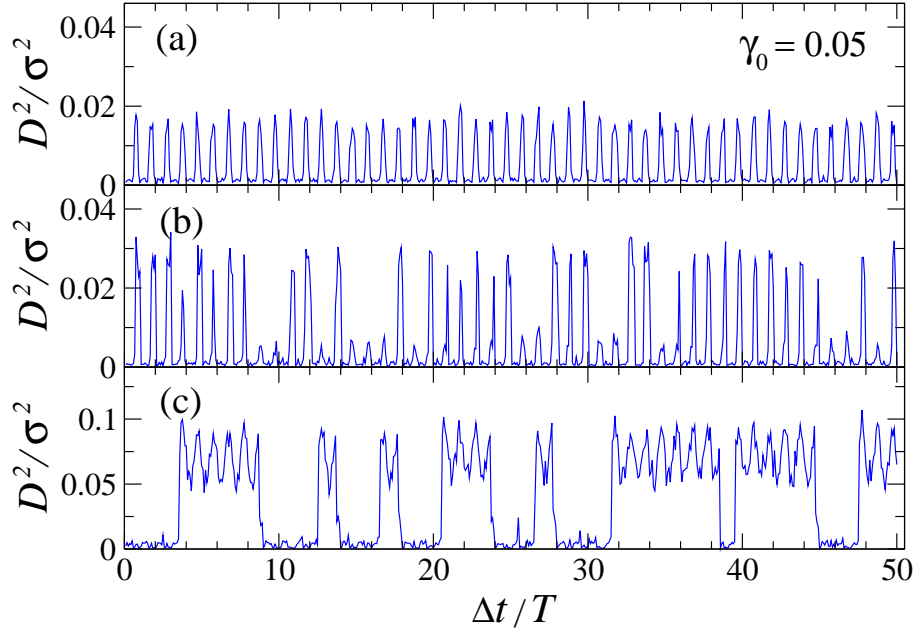}
\caption{(Color online) Variation of $D^2(0, \Delta t)$ as a
function of the time lag $\Delta t$ for three particles during $50$
periods at the strain amplitude $\gamma_{0}=0.05$ and oscillation
frequency $\omega\tau=0.001$.    Note that the vertical scale in the
panel (c) is different. }
\label{fig:d2min_time}
\end{figure}

Figure\,\ref{fig:d2min_time} shows the variation of the quantity
$D^2$ as a function of the time lag $\Delta t$ for selected
particles at the strain amplitude $\gamma_{0}=0.05$.    We compare
particle configurations separated by the time interval $\Delta t$
with respect to $t=0$.   It can be seen in
Fig.\,\ref{fig:d2min_time}\,(a) that the displacement of the
particle can be well described by the affine transformation during
the first half of each cycle, while it undergoes large nonaffine
displacements during the second half of each cycle.    Notice that
the function $D^2(0, \Delta t)$ is periodic with superimposed noise,
possibly because of the thermal fluctuations.   We find, however,
that local nonaffine displacements are not always periodic.   For
example, as shown in Fig.\,\ref{fig:d2min_time}\,(b), the particle
undergoes large nonaffine displacements separated by periods with
relatively small amplitudes of nonaffine displacements.  We comment
that this behavior is different from completely repetitive limit
cycles observed in athermal quasistatic
simulations~\cite{Reichhardt13,HernHoy13}.

\vskip 0.05in

\begin{figure}[t]
\includegraphics[width=12.cm,angle=0]{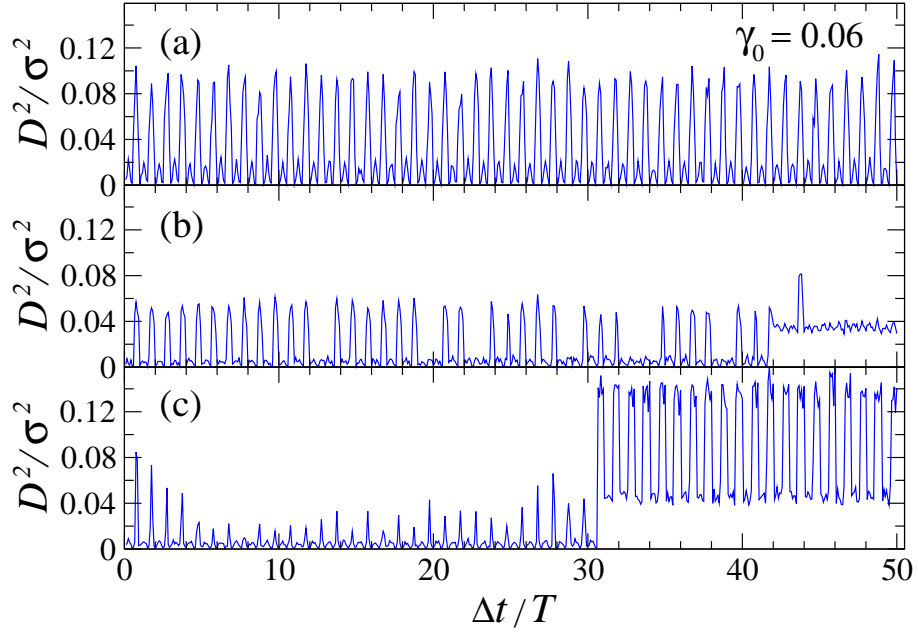}
\caption{(Color online) The dependence of the quantity $D^2(0,
\Delta t)$ as a function of the time lag $\Delta t$ for three
different particles at the strain amplitude $\gamma_{0}=0.06$ and
oscillation frequency $\omega\tau=0.001$.  }
\label{fig:d2min_time_gam06}
\end{figure}

Furthermore, in rare occasions, we detect particles that temporarily
escape the cage of its neighbors while still undergoing periodic
nonaffine displacements [see Fig.\,\ref{fig:d2min_time}\,(c)]. This
behavior is consistent with the observation of reversible cage jumps
during oscillatory shear deformation of a binary glass, where
particle trajectories were analyzed using the cage detection
algorithm~\cite{Priezjev13}.    It was also found that irreversible
cage jumps typically occur after $10^2-10^3$ cycles at the strain
amplitude $\gamma_{0}=0.05$ and frequency $\omega\tau=0.02$, when
the root mean square displacement of particles becomes comparable
with the cage size~\cite{Priezjev13}.     To remind, the critical
strain amplitude $\gamma_{0}=0.06$ marks the transition from a slow
dynamics with a broad subdiffusive plateau to a diffusive regime
governed by irreversible displacements of
particles~\cite{Priezjev13}. Finally, examples of repetitive
nonaffine displacements at the strain amplitude $\gamma_{0}=0.06$
are shown in Fig.\,\ref{fig:d2min_time_gam06} for three particles.
It can be seen that during $50$ oscillation periods, the
displacements are reversible except for the time lags $\Delta t
\approx 42\,T$ in Fig.\,\ref{fig:d2min_time_gam06}\,(b) and $\Delta
t \approx 31\,T$ in Fig.\,\ref{fig:d2min_time_gam06}\,(c), when
sudden irreversible rearrangements take place.

% the local particle configurations

\vskip 0.05in

% snapshots of clusters

\begin{figure}[t]
\includegraphics[width=12.cm,angle=0]{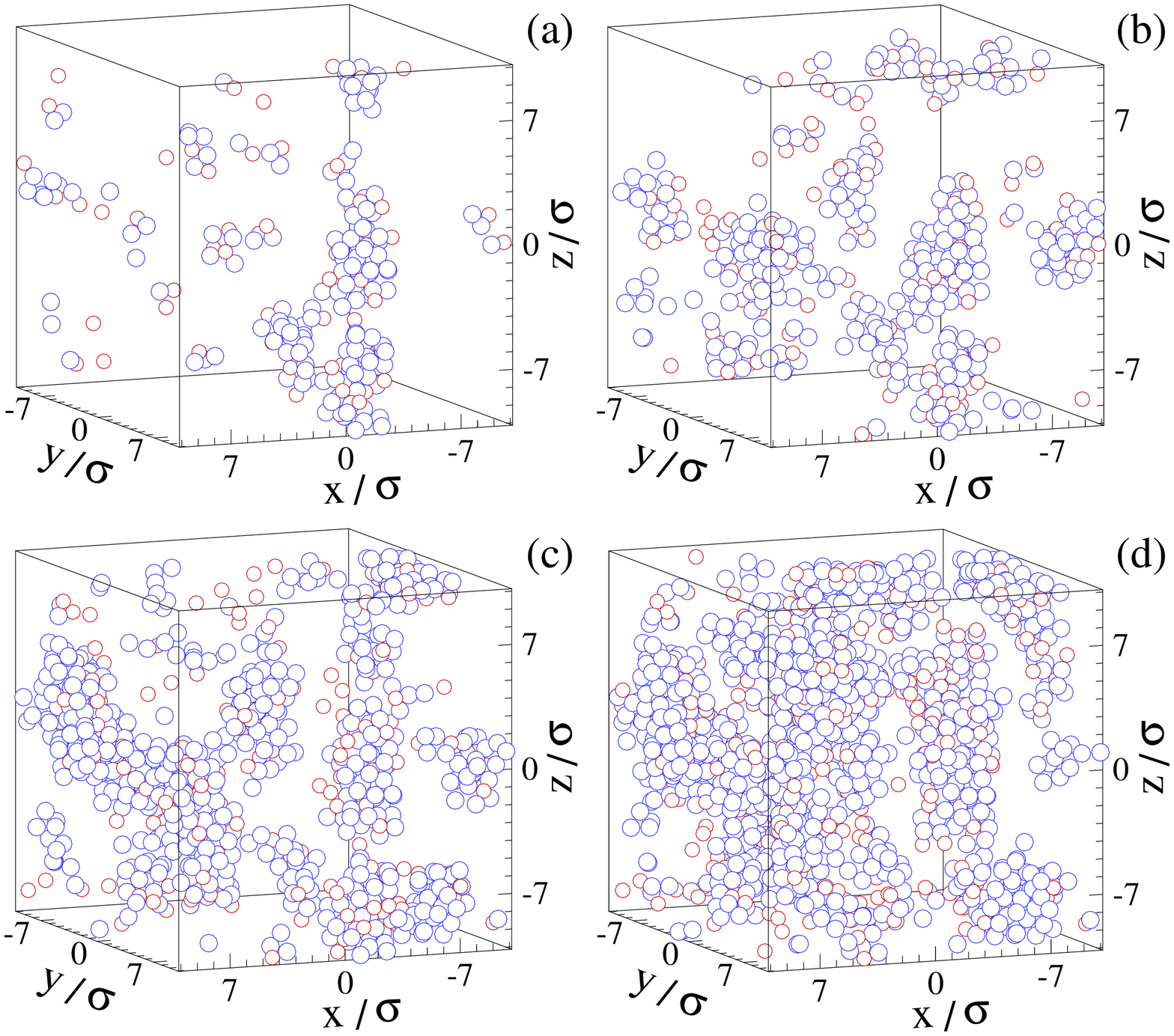}
\caption{(Color online) The positions of particles of type A (large
blue circles) and particles of type B (small red circles) at the
shear strain $\gamma(t=T/4)=\gamma_0$ and $D^2(0,
T/4)>0.01\,\sigma^2$ for the strain amplitudes (a)
$\gamma_{0}=0.02$, (b) $\gamma_{0}=0.03$, (c) $\gamma_{0}=0.04$, and
(d) $\gamma_{0}=0.05$.  }
\label{fig:snapshot_clusters}
\end{figure}

Typical configurations of particles with large nonaffine
displacements ($D^2>0.01\,\sigma^2$) after a quarter of a cycle are
presented in Fig.\,\ref{fig:snapshot_clusters} for the strain
amplitudes $0.02\leqslant\gamma_{0}\leqslant0.05$. It is evident
from Fig.\,\ref{fig:snapshot_clusters}\,(a) that even at the small
strain amplitude $\gamma_{0}=0.02$, these particles form relatively
large, compact clusters as well as a few isolated clusters that
consist only of a few atoms.    With increasing strain amplitude,
the cluster size increases and becomes comparable with the system
size at $\gamma_{0}\geqslant0.04$.    Thus, we identify a
\textit{dynamical process} that involves a correlated motion of
particles and leads to percolation of nonaffine displacements at
sufficiently large strain amplitudes.    Note that snapshots of
particles shown in Fig.\,\ref{fig:snapshot_clusters}\,(a)-(d) were
taken in the same sample, and, therefore, the clusters at different
strain amplitudes appear to be spatially correlated.    We also
mention that the cluster size distribution was not computed in our
study because of the limited statistics.   Since only a few large
clusters per sample are formed, and these clusters are nearly
reversible over fifty cycles, it is expected that a reliable cluster
size distribution can be obtained by averaging over a larger number
of independent samples (e.g., $10^3$ instead of $8$). Interestingly,
a power-law distribution of energy drops below yield strain was
recently reported in a model 2d solid subject to oscillatory
quasistatic shear~\cite{IdoNature15}.

\vskip 0.05in

% comments on average d2min

\begin{figure}[t]
\includegraphics[width=12.cm,angle=0]{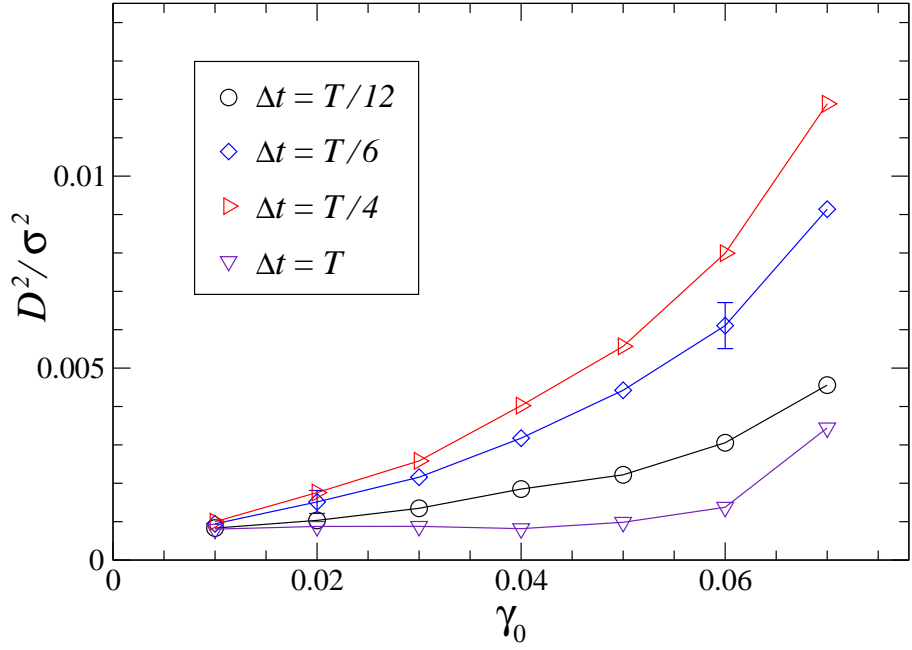}
\caption{(Color online) The averaged quantity $D^2(0, \Delta t)$ as
a function of the strain amplitude $\gamma_{0}$ for the time lags
$\Delta t = T/12$ ($\circ$), $T/6$ ($\diamond$), $T/4$
($\triangleright$), and $T$ ($\triangledown$), where $T$ is the
oscillation period.  The shear strain $\gamma(t)$ at $t = \Delta t$
is given by Eq.\,(\ref{Eq:strain}). }
\label{fig:ave_d2min_gam0}
\end{figure}

The dependence of the quantity $D^2(0, \Delta t)$ as a function of
the strain amplitude is plotted in Fig.\,\ref{fig:ave_d2min_gam0}
for different time lags $\Delta t$, which determine shear strain
according to Eq.\,(\ref{Eq:strain}).   The nonaffine displacements
were averaged over all particles in eight independent samples with
respect to $t=0$.    It appears that at the smallest strain
amplitude $\gamma_{0}=0.01$, the difference in $D^2(0, \Delta t)$
for various time lags in Fig.\,\ref{fig:ave_d2min_gam0} is barely
noticeable because only a small fraction of particles undergo large
nonaffine displacements (see Fig.\,\ref{fig:d2min_pdf}). As
expected, the average of $D^2(0, \Delta t)$ increases with
increasing shear strain $\gamma(\Delta t)$ and it becomes greater
than $0.01\,\sigma^2$ at the maximum strain
$\gamma(T/4)=\gamma_{0}=0.07$.   Note also that $D^2(0, T)$ after a
full back-and-forth cycle is not zero, which implies that relative
positions of a particle and its nearest neighbors are not exactly
reversible even at small strain amplitudes due to thermal motion of
particles within their cages.  This is consistent with the results
shown in Fig.\,\ref{fig:d2min_time}. Furthermore, the relatively
large increase of $D^2(0, T)$ at the strain amplitudes
$\gamma_{0}=0.06$ and $0.07$ in Fig.\,\ref{fig:ave_d2min_gam0}
reflects the occurrence of irreversible displacements after a single
cycle.    These results confirm that with increasing strain
amplitude above $\gamma_{0} = 0.06$, the number of irreversible cage
jumps after one cycle increases and the root mean square
displacement of particles becomes greater than the cage
size~\cite{Priezjev13}.    These conclusions also agree with the
definition of the yielding transition that occurs at the largest
strain amplitude below which the microstructure is
reversible~\cite{Arratia13}.

% 10 T in last fig?

\section{Conclusions}

In summary, molecular dynamics simulations were conducted to study
structural relaxation in a three-dimensional model glass submitted
to periodic shear.   We considered the Kob-Andersen binary mixture
at the temperature which is well below the glass transition.   The
nonaffine component of the particle displacement was evaluated
during multiple time intervals with respect to the system
configuration at zero strain.     It was found that even at strain
amplitudes below yield, some particles undergo large nonaffine
displacements that are reversible after each cycle.   The magnitudes
of the nonaffine displacement, computed after a quarter of a cycle,
are approximately power-law distributed with the slope that depends
on the strain amplitude.    During cyclic loading, mobile particles
tend to aggregate into transient clusters which become larger with
increasing strain amplitude. The probability of irreversible
rearrangements after a full cycle also increases, leading to
permanent structural relaxation of the material at large strain
amplitudes.

\section*{Acknowledgments}

Financial support from the National Science Foundation (CNS-1531923)
is gratefully acknowledged.    The author would like to thank I.
Regev for useful comments.  Computational work in support of this
research was performed at Michigan State University's High
Performance Computing Facility and the Ohio Supercomputer Center.

%%%%%%%%%%%%%%% FIGURES %%%%%%%%%%%%%%%%%%%%%%%

% snapshot of the system

% D2min pdf during 10T various

% individual particles: d2min time gamma0=0.05

% individual particles: d2min time gamma0=0.06

% typical clusters of reversible jumps (when D2 large)

% average d2min versus gamma_zero

\bibliographystyle{prsty}

\end{document}